\documentclass[iop]{emulateapj}

\usepackage{epsfig}
\usepackage{amsmath}
\usepackage{natbib}
\usepackage{apjfonts}

\slugcomment{accepted in ApJL, 2011}

\shorttitle{Polarization of the Charge-Exchange X-rays Induced in the Heliosphere}
\shortauthors{Gacesa, M\"uller, C\^ot\'{e}, Kharchenko}

\begin{document}


\title{Polarization of the Charge-Exchange X-rays Induced in the Heliosphere}



\author{M. Gacesa\altaffilmark{1,2}, H.-R. M\"uller\altaffilmark{3}, R. C\^ot\'{e}\altaffilmark{2}, and V. Kharchenko\altaffilmark{1,2}}

\email{mgacesa@cfa.harvard.edu}


\altaffiltext{1}{Harvard-Smithsonian Center for Astrophysics,
   60 Garden Street, Cambridge, MA 02138, USA}
\altaffiltext{2}{Physics Department, University of Connecticut,
   2152 Hillside Road, Storrs, CT 06269, USA}
\altaffiltext{3}{Department of Physics and Astronomy, Dartmouth College,
   6127 Wilder Laboratory, Hanover, NH 03755, USA}

\begin{abstract}
We report results of a theoretical investigation of polarization of the X-ray emissions induced in charge-exchange collisions of fully stripped solar wind ions C$^{6+}$ and O$^{8+}$ with the heliospheric hydrogen atoms. 
The polarization of X-ray emissions has been computed for line-of-sight observations within the ecliptic plane as a function of solar wind ion velocities, including a range of velocities corresponding to the slow and fast solar wind, and Coronal Mass Ejections. 
To determine the variability of polarization of heliospheric X-ray emissions, the polarization has been computed for solar minimum conditions with self-consistent parameters of the solar wind plasma and heliospheric
gas and compared with the polarization calculated for an averaged solar activity.
We predict the polarization of charge-exchange X-rays to be between 3\% and 8\%, depending on the line-of-sight geometry, solar wind ion velocity, and the selected emission lines.
\end{abstract}


\keywords{atomic processes - interplanetary medium - X-rays: diffuse background -
   X-rays: polarization}



\section{Introduction}
Charge-exchange (CX) collisions between highly charged Solar Wind (SW) ions and neutral gas were recently identified as an efficient mechanism for production of EUV and soft X-ray emissions 
\citep{2001SSRv...97..401R,2006A&A...460..289K,2009SSRv..143..217K,2007P&SS...55.1135B,2009Snowden}. There are indications that the CX collisions in the Heliosphere and Geocorona yield between 50\% and 80\% of the observed soft (below 1 keV) X-ray photons, making them a significant contributor to the soft X-ray background.
Heliospheric CX X-ray emissions are sensitive to the parameters of the SW plasma and strong correlations between variations in the SW intensity and composition, and intensity of the soft X-ray background were observed and analyzed \citep{2001SSRv...97..401R,2008AGUFMSH21B1596R,2009SSRv..143..217K}. 
These findings indicate that the heliospheric CX X-rays could be used for diagnostics of the solar wind composition and velocities, as well as an independent probe of spatial distribution of the heliospheric neutral gas.
The X-ray emission, if measured, could provide additional insight into interaction of the SW plasma with the neutral heliospheric gas. Parameters of this interaction are critical for modeling of the interstellar gas distribution and for interpretation of observational data on the diffuse soft X-ray emission from the Local Bubble \citep{2005Sci...307.1447L,2008ApJ...676..335H,2009ApSS.323....1W}.

While polarization of optical emissions is readily used in investigation of various astrophysical objects, detection of X-ray polarization remains a technically challenging feat. Currently available observational data do not include polarization. However, the angular distribution of polarized emissions is known to be anisotropic \citep{1973RvMP...45..553F,2000JPhB...33.5091T}. If the CX X-ray emissions are strongly polarized, the total intensities measured in current satellite observations should be redefined to adjust for the anisotropy.
It is realistic to expect that X-ray polarization will be investigated in future space missions, making an accurate theoretical consideration of these phenomena necessary. Currently, polarization data are available from laboratory experiments for some CX collisions of atoms and ions of astrophysical interest. For example, polarization spectroscopy of ${\rm O}^{5+} (1s^2 3p)$ produced in collisions of ${\rm O}^{6+}$ with He and ${\rm H}_2$, for projectile velocities from 740 to 1200 km s$^{-1}$, showed that the CX X-rays are polarized and strongly dependent on the projectile velocity \citep{2000JPhB...33.5091T}. 
We expect CX X-ray emissions produced in astrophysical environments, such as Jupiter heavy ion auroras \citep{2006GeoRL..3311105K,2008JGRA..11308229K} and coronal mass ejections, to exhibit similar properties.

In this Letter, we report the results of our theoretical study of polarization of the X-rays emitted in charge-exchange collisions between fully stripped SW ions and neutral heliospheric hydrogen. In particular, we investigate the dependence of the X-ray polarization on distribution of the SW plasma, heliospheric neutral gas and SW ion velocity.

\section{Polarization of the emissions induced in charge-exchange collisions}

Descriptions of a theoretical model of charge-exchange-induced X-rays in the Solar System and calculated emission spectra have been previously published  \citep{1998JGR...10326687K,2008JGRA..11308229K,2000JGR...10518351K,2006A&A...460..289K,2007P&SS...55.1135B}. We consider collisions between heavy SW ions and hydrogen gas, ${\rm X}^{Q+} + {\rm H} \rightarrow {\rm X}^{*(Q-1)+} + {\rm H}^{+}$, taking into account that H is a predominant component of interstellar gas. This model can be easily extended to include minor components of the interstellar gas such as He.

The line-of-sight (LOS) intensity, expressed as number of photons per square centimeter per second, of an emission line of the wavelength $\lambda$ is given by
\begin{equation}
   \label{eq:R}
   I_{\lambda}^{\rm LOS} = \frac{1}{4\pi} \int_{R_d}^{R_h} N_{{\rm X}^{Q+}}(r) n_{\rm H}(r) v_{\rm rel} \sigma_{{\rm H},{\rm X}^{Q^+}} dr ,
\end{equation}
where $N_{{\rm X}^{Q+}}(r)$ is the density of solar wind ions, $n_{\rm H}(r)$ is the density of hydrogen, $v_{\rm rel}$ is the relative collision velocity, $\sigma_{{\rm H},{\rm X}^{Q^+}}$ is the electron capture cross section for an excitation of the X-ray emission with the wavelength $\lambda$. In Eq. (\ref{eq:R}), $R_{\rm d}$ is the position of the X-ray detector, and $R_{\rm h}$ is the distance to the heliopause, as defined for a particular geometry and distribution of heliospheric plasma. For simplicity, we assume that $v_{\rm rel} \approx v$, where $v$ is the velocity of SW ions. The polarization of the heliospheric X-ray emission in LOS observations can be defined as $P=(I_{\parallel}-I_{\perp})/(I_{\parallel}+I_{\perp})$, where $I_{\parallel}$ and $I_{\perp}$ are the detected intensities, defined below.

Projections of angular momenta of the electronic states of the excited SW ions produced in CX collisions are oriented along the local velocity direction of the SW plasma. Assuming that a satellite detector can analyze linear polarization of X-ray emissions, the intensity $I(\psi)$ of the radiation, emitted from a selected small volume of the heliosphere and normalized to a single ion emission, can be expressed as \citep{1973RvMP...45..553F}
\begin{equation}
   \label{eq:I_simple}
   I(\psi) = \frac{n_{\rm H} v \sigma_{{\rm H},{\rm X}^{Q^+}}}{4\pi R^2} \left(1-\frac{1}{2} h^{(2)} A_0^{\rm det} + \frac{3}{2} h^{(2)} A_{2+}^{\rm det} (\psi) \right) ,
\end{equation}
where $R$ is the distance between the detector and the CX event, and $h^{(2)}$ is the ratio of recoupling coefficients which contain the elements of orthogonal transformations between the initial and final quantum state \citep{1973RvMP...45..553F}. The angle $\psi$ indicates the orientation of the linear polarization analyzer. In the assumed geometry, we define the polarization parallel and perpendicular to the ecliptic plane as $I_{\parallel} = I(\psi=0)$ and $I_{\perp} = I(\psi = \pi/2)$.
Here, $A_0^{\rm det}$ and $A_{2+}^{\rm det}$ are alignment parameters
\begin{eqnarray}
  A_0^{\rm det} & = & \frac{1}{2} A_0^{\rm col} (3\cos^2{\theta}-1), \nonumber \\
  A_{2+}^{\rm det}(\psi) & = & \frac{1}{2} A_0^{\rm col} \sin^2{\theta} \cos{2\psi} ,
\end{eqnarray}
where $A_0^{\rm col}$ is the alignment tensor, $\theta$ is the angle between the ion propagation direction and the detection axis.
For the defined geometry, the alignment tensor $A_0^{\rm col}$ is a scalar which can be expressed in terms of the partial cross section $\sigma(m_i,v)$ for electron capture into the atomic state $\lvert j_i m_i \rangle$:
\begin{equation}
  A_0^{\rm col} = \frac{ \sum_{m_i} \left[ 3 m_i^2 - j_i(j_i+1) \right] \sigma(m_i,v) }
                   { j_i(j_i+1)\sum_{m_i}{\sigma(m_i,v)} } .
\end{equation}
The parallel and perpendicular intensities in the detector frame are then
\begin{eqnarray}
  I_{\parallel} & = & \frac{n_{\rm H} v \sigma_{{\rm H},{\rm X}^{Q^+}}}{8\pi R^2} \left[1-\frac{1}{2} h^{(2)} A_0^{\rm col} (1-3\sin^2{\theta})\right], \\
  I_{\perp} & = & \frac{n_{\rm H} v \sigma_{{\rm H},{\rm X}^{Q^+}}}{8\pi R^2} \left( 1-\frac{1}{2} h^{(2)} A_0^{\rm col} \right) ,
  \label{eq:intdet}
\end{eqnarray}
Since the cross sections $\sigma(m_i,v)$ depend on the relative velocity of collisions $v$ \citep{1998ADNDT..68..279H}, the polarization is also velocity dependent.

In order to calculate the X-ray polarization along a LOS, we adopt planar geometry with Earth (the X-ray detector is assumed to be on a satellite in Earth's orbit), Sun and the CX event located in the ecliptic plane and inside the heliosphere. We consider Earth to be approximately at the Fall point ($R_0=1$ AU) and specify the LOS direction by the angle $\alpha$, such that for $\alpha=0$ the LOS points toward Sun, for $\alpha=\pi/2$ in the direction opposite to the interstellar flow (upwind), and for $\alpha=-\pi/2$ along the interstellar flow (downwind). Here, $R$ and $r$ are the distances from Earth (detector) and Sun to the CX event, respectively.

We first illustrate general features and estimate the X-ray polarization using simplified heliospheric distributions of the SW plasma and neutral gas. If the heliospheric hydrogen density is assumed to be constant, $n_{\rm H}(r) = n_{\rm H}^{(0)}$, and distribution of SW ions to be isotropic as seen from the Sun's location, $N_{{\rm X}^{Q+}}(r) = N_{{\rm X}^{Q+}}^{(0)} \left(R_0 / r\right)^2$, where $N_{{\rm X}^{Q+}}^{(0)}$ is the ion density at 1 AU, we can analytically integrate Eqs. (\ref{eq:intdet}) up to the heliopause, which we assume to be a sphere of the radius $R_{{\rm h}}$. 

Resulting expressions for the LOS angle $0<\alpha<\pi$ are 
\begin{eqnarray}
  I_{\perp}^{{\rm LOS}} & = & \frac{D}{\sin \alpha} \left( 1 - \frac{h^{(2)} A_0^{\rm col}}{2} \right) \left( \pi - \alpha - \beta  \right) \\
  I_{\parallel}^{{\rm LOS}} & = & I_{\perp}^{{\rm LOS}} + \frac{3 D h^{(2)} A_0^{\rm col}}{4 \sin \alpha} 
  \left( \pi - \alpha - \beta + \frac{\sin 2\alpha + \sin 2\beta}{2} \right)
  \label{eq:Pana}
\end{eqnarray}
where $D = {N_{{\rm X}^{Q+}}^{(0)} n_{\rm H} v \sigma_{{\rm H},{\rm X}^{Q^+}} R_0}/{8 \pi}$, and $\beta = \arcsin{\left[ (R_0 / R_h) \sin \alpha \right]}$. The X-ray intensities are given in cm$^{-2}$ s$^{-1}$. 
The simplified analytic model was used for comparison with numerical solutions obtained in the next section. It overestimates the X-ray intensities and degree of polarization by about a factor three when compared to more realistic gas and plasma distributions used in the next section.

\section{Polarization for realistic distributions of the heliospheric gas and plasma}

\begin{figure*}[htbp]
\begin{center}
\includegraphics[width=5.7cm]{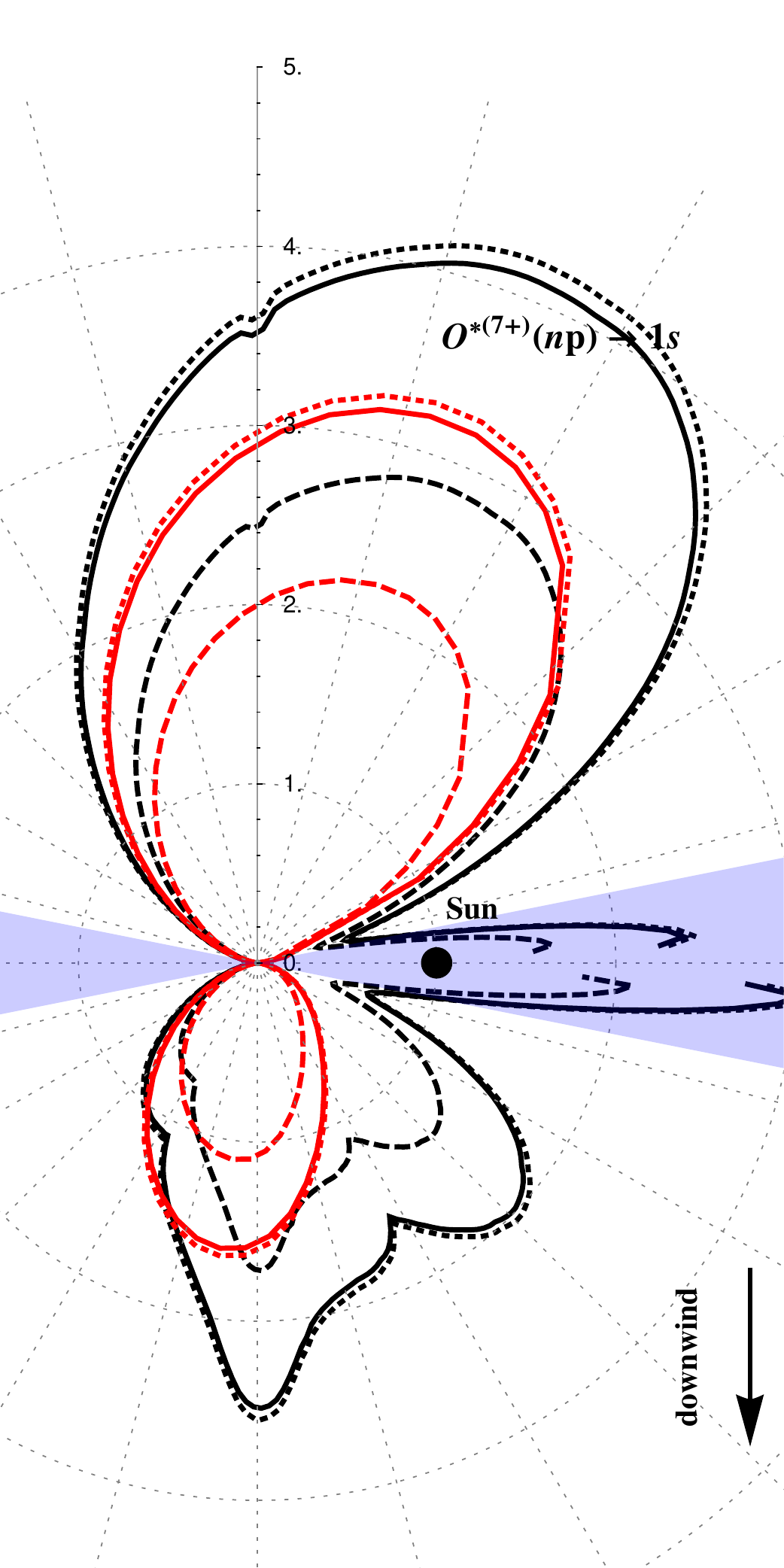}
\hspace{1.5cm}
\includegraphics[width=6.5cm]{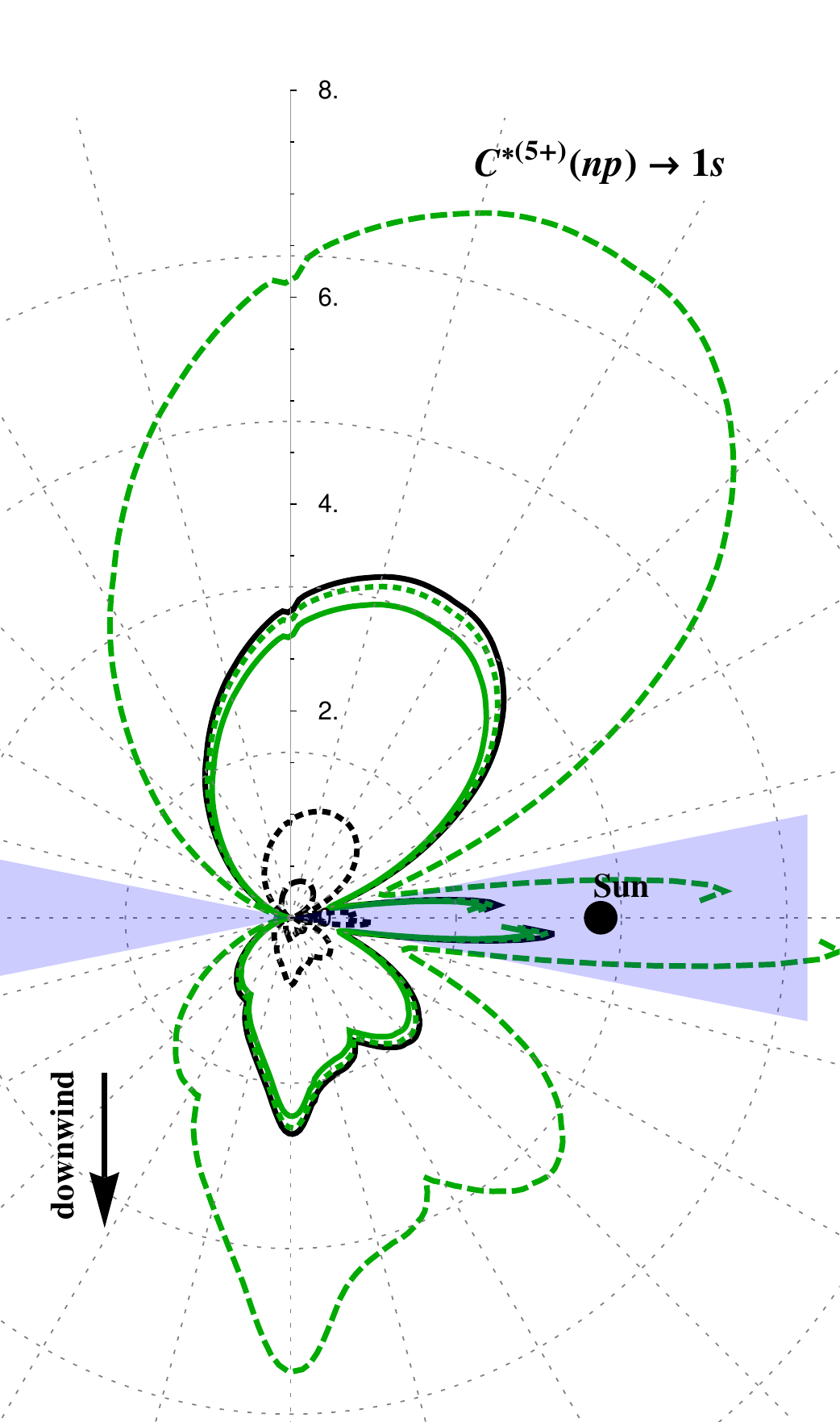}
\caption{Angular dependence of the polarization of CX X-rays emitted in a single de-excitation from the 4$p$ (solid lines), 5$p$ (dashed lines) and 6$p$ (dotted lines) excited states. Polarization for the angle of observation $\alpha$ is expressed as the radius-vector $P(\alpha)$ (\%). The axis for $\alpha = \pi/2$ (LOS in the upwind direction) is indicated, and the angle $\alpha$ increases counterclockwise. Blue cones indicate regions within $\pm 10^{\circ}$ of the Sun (right side) or directly opposite (left). Left panel shows polarization of the X-rays emitted by O$^{*(7+)}$, calculated for the slow SW using the CHM (red lines) and MHD (black lines) models. Right panel shows polarization for the CX emissions from C$^{*(5+)}$ for the fast (green lines) and slow (black lines) SW calculated within the MHD model.
}
\label{f:polarpl}
\end{center}
\end{figure*}

We calculated polarization of X-rays for single-photon relaxation from the $4p$, $5p$ and $6p$ excited states of C$^{*(5+)}$ and O$^{*(7+)}$ produced in CX collisions of C$^{6+}$ and O$^{8+}$ ions with heliospheric hydrogen. Probability to populate these states is an order of magnitude or more higher than the probability to populate other excited states \citep{1998ADNDT..68..279H}, and the selected ions characterize the slow SW, which dominates the ecliptic plane \citep{2006A&A...460..289K,2009ApJ...697.1214K}. Radiative cascade from the excited states was not included in the calculation since it does not contribute significantly to populations of considered excited states. Electron capture cross sections depend on the relative velocity of colliding particles; we computed the polarization for solar wind ion velocities from 200-2200 ${\rm km~s^{-1}}$, using the cross sections published by \citet{1998ADNDT..68..279H}. 
The upper limit of the considered velocity range corresponds to ion velocities found in the fastest Coronal Mass Ejections (CMEs) \citep{SOHO_LASCO_Catalog}, and the selected ion velocity range includes both the slow (400 ${\rm km~s^{-1}}$) and fast (750 ${\rm km~s^{-1}}$) SW \citep{2003Sci...302.1165S}.

Two different models of the heliospheric plasma were used in our calculation. The classical ``hot model'' (CHM) \citep{2004A&A...418..143L,2005Sci...307.1447L} assumes simple trajectories in a hot environment and it is appropriate for describing density distributions of hydrogen and helium averaged over short-term solar activity, including solar minima and maxima. This model has been employed in calculations of the heliospheric CX X-ray background averaged on the solar activity cycle \citep{2004ApJ...617.1347P}. The CHM model does not include the termination shock and the distribution of charged SW particles was assumed to be isotropic with respect to the Sun and to follow $N_{{\rm X}^{Q+}}(r) = N_{{\rm X}^{Q+}}^{(0)} \left(R_0/r \right)^2$. Since the heliopause is not defined within the CHM model, the integration limit was set to 300 AU.

The second model \citep{Mueller06} is based on a MHD multifluid calculation with four interpenetrating fluids, where one `fluid' represents protons of the interstellar and SW plasma, while three remaining components are used to model the neutrals (see \citet{Zank96} for details). We used heliospheric hydrogen density distributions from the MHD model, and density distributions of C$^{6+}$ and O$^{8+}$ ions were obtained by scaling the proton density from the model according to ion abundances of the fast and slow SW \citep{2000ApJ...544..558S,2006A&A...460..289K}. The intensities were integrated up to the heliopause and plasma density variation across the termination shock was included in the calculation.

For both models, we considered solar conditions corresponding to a typical solar minimum. All external perturbations that could mix populations of ionic excited states, such as heliospheric and interstellar magnetic fields, were neglected. The intensities were calculated in the entire ecliptic plane.

\section{Results}

\begin{figure*}[ht]
\begin{center}
\includegraphics[width=0.9\textwidth]{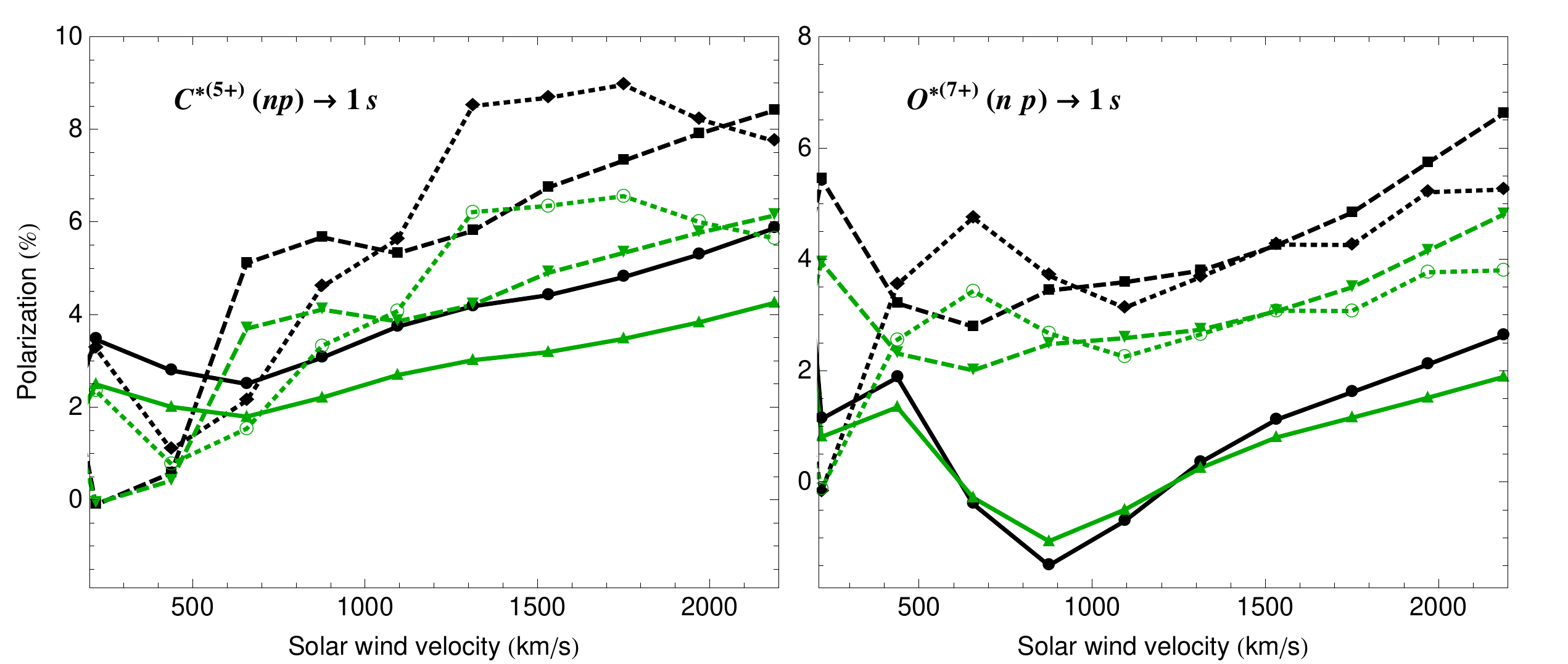}
\caption{Dependence of CX X-ray polarization on ion velocity for C$^{*(5+)}$ and O$^{*(7+)}$. Emissions from the $4p$ (solid), $5p$ (dashed) and $6p$ (dotted) excited states for LOS in the upwind (black lines) and downwind (green lines) directions are shown.}
\label{f:PvC6O8}
\end{center}
\end{figure*}

We calculated polarization maps of CX X-ray emissions in the ecliptic plane for C$^{*(5+)}$ and O$^{*(7+)}$ ions (Figure \ref{f:polarpl}). 
The angular dependence of the polarization is a result of interplay of several factors. The parallel component of the intensity, $I_{\parallel}$, depends on the angle of observation according to the factor $(1-3 \sin^2 \theta)$ (in Eq. (\ref{eq:intdet})), while the total intensity is dependent on the density of heliospheric plasma, which is rather complex \citep{Mueller06}. 
Namely, the ion density distribution is the greatest outside of the Solar corona and diminishes with the distance, roughly following the inverse square law inside of the termination shock. 
Neutral H is the dominant component of the interstellar flow. It penetrates into the heliosphere from $(\lambda, \beta)_{\rm HE} = (252.3^{\circ},8.5^{\circ})$ direction in the helio-ecliptic coordinates \citep{2005Sci...307.1447L}, where H atoms are ionized as they encounter the solar wind and radiation. Consequently, the H gas density decreases as the ionization rate increases in the vicinity of Sun. The resulting H distribution is anisotropic and sensitive to the level of solar activity.
Thus, the CX X-ray polarization map is affected by both the geometrical factor and the anisotropicity of the plasma density.

We first computed the polarization maps of CX X-ray emissions from the $4p$, $5p$ and $6p$ states of O$^{*(7+)}$ induced by the slow SW, where we used the CHB and MHD models to describe the densities of H and ions for conditions typical of a solar minimum (Figure \ref{f:polarpl}, left panel). 
Both models predict the highest polarization in the upwind direction of observation that vanishes as the LOS angle approaches $\pi$. Ratios of polarization calculated for different initial states remain the same within the models, although the values calculated using the MHD model are about a quarter higher.
The highest polarization is predicted for $\alpha=68.3^{\circ}$, and $\alpha=71^{\circ}$, for the MHD and CHM model, respectively. 
Quantitatively, polarization of the X-ray emissions from the $4p$ and $6p$ states in the MHD model is about $P=4\%$, and $P<3\%$ for the initial $5p$ state. 
For comparison, the simple analytic model given in Eq. (\ref{eq:Pana}), predicts $P=13.4$ \% in the LOS direction $\alpha=51.4^{\circ}$ for emissions induced from the $4p$ state by the slow SW, and scales similarly to the MHD model for the $5p$ and $6p$ states.


To better illustrate the dependence of the CX-induced X-ray polarization on the velocity, we show an analogous polarization map for the slow and fast SW (Figure \ref{f:polarpl}, right panel) constructed for the MHD model.
We predict lower polarization of the CX X-rays induced by the fast SW than by the slow SW. The only exception are CX emissions from C$^{*(5+)}(5p)$, which we find to have the polarization of about 7\% for the optimal angle of observation, $\alpha \approx 70^{\circ}$.
Although the fast SW is not present at low (below $\beta_{\rm HE} = \pm 20^{\circ}$) heliographic latitudes, the calculated velocity-dependence of polarization may give additional insight in understanding the CX emissions induced at higher heliographic latitudes, as well as in similar astrophysical environments. 

In Figure \ref{f:PvC6O8} we illustrate the polarization for a broader range of ion velocities for the three excited states of C$^{*(5+)}$ and O$^{*(7+)}$ in the upwind ($\alpha=90^{\circ}$) and downwind ($\alpha=270^{\circ}$) directions. 
The polarization for the optimal LOS, $\alpha=68.3^{\circ}$, is about 17\% higher than in the upwind direction. As an overall trend, the polarization increases as the ion velocity increases for ion velocities higher than 1000 km s$^{-1}$, except for the $6p$ state in C$^{*(5+)}$ where the polarization decreases for velocities higher than 1700 km s$^{-1}$. 
This behavior is a reflection of dependence of the CX partial cross sections on velocity. While for high velocities we expect the populations of different magnetic sublevels to asymptotically approach their statistical values, SW ion velocities are not high enough to clearly show that trend.

\section{Conclusion and future directions}

In this work, we present the first calculations of polarization of the heliospheric X-rays induced by the CX collisions with fully stripped oxygen and carbon ions, O$^{8+}$ and C$^{6+}$.
To analyze its velocity and directional dependence, we calculated the polarization in the ecliptic plane as a function of the LOS for ion velocities from 200-2200 km s$^{-1}$, where the heliospheric plasma was described using two models of different complexity \citep{2004A&A...418..143L,2000JGR...10527419M}. Our calculations indicate that the CX heliospheric X-rays are mildly polarized and that the polarization depends on the SW ion velocity. 

While this study was restricted to C$^{6+}$ and O$^{8+}$ ions in a simplified geometry, it nevertheless, illustrates a rather general property of the CX-induced radiation. In the considered interval of ion velocities colliding ion and atom form an intermediate  quasi-molecule whose projection of electronic angular momentum is quantized along the  direction of the quasi-molecular axis. In collisions, quasi-molecular axes of different ion-atom pairs rotate until, at the end of encounter, they are oriented along the ion velocity vector. Thus, CX collisions yield an ensemble of aligned excited ions that emit polarized photons. The polarization reflects a directional orientation of local collisional velocities in the regions of the SW plasma that are mostly responsible for production of the X-ray flux.
Analogously, we may expect a relatively high level of polarization of the CX X-ray emissions from cometary atmospheres or in Jovian polar X-ray auroras, where the ion velocity co-orientation in regions of the X-ray-producing CX collisions may be very high. Fast CMEs, propagating in the interplanetary gas, could also produce highly polarized CX X-ray emissions, particularly if the X-ray detector is located close to the CME's plasma stream.

Polarization measurements could be used as a powerful tool to supplement the CX spectra and provide additional insight into underlying astrophysical processes. An example of such an environment, directly related to this work, is the Local Bubble. Several recent studies backed by new observational data questioned the validity of the current Local Bubble picture, arguing that the contribution of heliospheric X-rays to the soft X-ray background may be higher than previously thought \citep{2001SSRv...97..401R,2007A&A...475..901K,2008AGUFMSH21B1596R,2008ApJ...676..335H,2009ApSS.323....1W}. The polarization measurements could help identify the contribution from the heliospheric CX radiation in the diffuse X-ray background, especially if performed for different levels of solar activity.

As a continuation of this work, a detailed map of the polarization of the SW induced X-rays in the Solar System should be constructed using results of an elaborate 3D MHD model of the heliospheric plasma and neutral gas. If computed for different solar conditions, such maps could be used as an independent method for determining velocity and spatial distributions of the solar wind plasma.
Finally, the investigation of CX-induced X-ray polarization may be extended to other interesting topics, including the He focusing cone, dependence of polarization on the solar activity, or regions of the SW plasma turbulence \citep{2008ApJ...682.1404B}.

VK acknowledges support by NASA grant NNX08AH51G and HRM acknowledges support by NSF grant AST-0607641.


%

\end{document}